\documentstyle[aps,prl,multicol]{revtex}
\begin{document}

\title{Phase separation and crystallisation of polydisperse
hard spheres}

\author{{\bf Richard P. Sear\footnote{{
%\large
Address from September 1998: Department of Physics,
University of Surrey, Guildford, Surrey GU2 5XH, United Kingdom}}}\\
~\\
Department of Chemistry and Biochemistry\\
University of California, Los Angeles\\
Los Angeles, California 90095-1569, U.S.A.\\
email: sear@chem.ucla.edu}

\date{\today}

\maketitle

\begin{abstract}
Hard spheres with a polydispersity above approximately 8\% are shown to
crystallise into two phase-separated solid phases.
A polydispersity above 8\% is too large to be tolerated by a single
solid phase but phase separation produces two fractions with
polydispersities sufficiently narrow to allow them to crystallise.
It may not be possible to observe this in experiment due to the intervention
of a glass transition.
\end{abstract}

\vspace{0.2in}

%82.70.Dd Colloids
%64.70.Dv Solid-liquid transitions
%81.30.-t Phase diagrams and microstructures developed by solidification 
%and solid-solid phase transformations

%\newpage
\begin{multicols}{2}
\narrowtext

Experiments on synthetic colloids are inevitably on polydisperse
colloids. The colloidal particles are not all the same, some are
larger than average, some are smaller \cite{pusey91}.
This has prompted considerable theoretical study of the phase behaviour of
polydisperse systems, in particular the fluid--solid transition
of polydisperse hard spheres
\cite{barrat86,mcrae88,bolhuis96,bartlett97,bartlett}.
The conclusion of these studies is that a
solid of hard spheres can tolerate
a polydispersity of a little less than 0.1, beyond this it is never
stable with respect to the fluid phase.
The 0.1 refers to the ratio of the standard deviation to the mean,
denoted below by $s$. As Pusey \cite{pusey87} has noted,
this makes sense if we note that the lattice
constant of the solid is at most 10\% larger than the
diameter of the hard spheres. Hard spheres melt at a fraction
$\simeq 0.74$ of close packing \cite{hoover68} which corresponds
to a lattice constant $\simeq 1.11$ times the hard sphere diameter.
Only if the polydispersity is sufficiently narrow that
few or no spheres are larger than this lattice spacing can the mixture
crystallise \cite{pusey87}. We have derived an analytical free
energy for the solid phase of polydisperse hard spheres and
our results agree with those of previous authors
\cite{barrat86,mcrae88,bolhuis96,bartlett97,bartlett}.
We find that the Helmholtz free energy of the
solid phase of polydisperse hard spheres with $s>0.082$ is never
lower than that of the fluid phase at the same density.

However, all this takes no account of possible demixing. If
the spheres phase separate into fractions then these fractions will have
narrower polydispersities than the original mixture and so can crystallise.
Of course, this phase separation costs
some ideal mixing entropy but at high density spheres pack
more efficiently in the solid phase than in the fluid resulting
in a larger entropy in the solid.
The possibility of demixing has been mentioned before
\cite{barrat86,pusey87} but no calculations have been performed.
Here, calculations show polydisperse hard spheres
crystallising into two solid phases,
one containing the larger spheres and one containing the smaller spheres.
There is no fluid--fluid phase separation at the moderate
polydispersities we consider,
although there may well be at much larger polydispersities
\cite{cuesta,warren}. By moderate we mean polydispersities up to about
20\%.

The crystallisation of polydisperse hard-sphere-like colloids
into more than one solid phase
has not been observed experimentally. This is despite the
fact that it is almost certainly the equilibrium
behaviour. The discrepancy between theory and experiment may be due
to the intervention of the glass transition
\cite{vanmegen91,underwood94}. The cost
in demixing entropy and the less efficient packing of polydisperse
spheres into a solid vis-a-vis monodisperse spheres both
act to raise the coexisting densities of the fluid--solid transition.
It is quite possible that when the polydispersity is sufficiently
large for the equilibrium behaviour to be crystallisation into
two solid phases, the transition occurs at densities above
the glass transition and is thus impossible to observe in experiment.
Already for near-hard-sphere colloids with a polydispersity $s=0.075$
the fluid--solid and glass transitions appear to be very close \cite{pusey91}.

In a polydisperse mixture of spheres, spheres with a
range of diameters are present \cite{salacuse82,gualtieri82}; there is
a continuous distribution of sizes of spheres
with a number
density $\rho x(\sigma){\rm d}\sigma$ of spheres of size $\sigma$.
$\rho$ is the total number density of spheres.
The width of the polydispersity is characterised by a width parameter $w$.
The larger $w$ is the broader the distribution of sizes present in
the mixture. In the limit $w\rightarrow0$ we recover a monodisperse
system.
The functional form of the polydispersity should not matter too much;
we select a simple form, the hat function:
\begin{equation}
x(\sigma)=\left\{
\begin{array}{ll}
0 & ~~~~~~ \sigma< {\overline\sigma}-w/2 \\
w^{-1} & ~~~~~~ {\overline\sigma}-w/2\le \sigma\le {\overline\sigma}+w/2 \\
0 & ~~~~~~ \sigma > {\overline\sigma}+w/2 \\
\end{array}\right. ,
\label{xs}
\end{equation}
where ${\overline\sigma}$ is the mean diameter.
The standard deviation divided by its mean, of the distribution $x(\sigma)$
is $s=w/(\sqrt{12}\sigma)$.

An approximation to the Helmholtz free energy of the solid phase
may be obtained using a cell theory \cite{buehler51}.
Our system is athermal so the free energy is equal to minus
the entropy (ignoring the contribution from the momenta). Within
a cell theory the entropy per particle
is just the logarithm of the configuration
space available to one particle trapped in a cell formed by its
neighbours. These neighbouring particles are
taken to be fixed at their lattice positions,
which are the lattice constant $a$ from the centre of the cell.
For a solid of monodisperse spheres of diameter $\sigma$
the hard-core
repulsion prevents the particle from moving more than $(a-\sigma)$
from its lattice position in the direction of one
of its neighbours.
Then we can approximate the
cell by a sphere of radius $a-\sigma$ \cite{buehler51} and so obtain
a free energy per particle $f_s$ in the solid phase
\begin{equation}
\frac{f_s}{kT}=-\ln\left[\frac{4\pi}{3}(a-\sigma)^3\right],
\label{asmono}
\end{equation}
where for a close-packed lattice, $a=(\eta_{cp}/\eta)^{1/3}$;
$k$ is Boltzmann's constant and $T$ is the temperature.
The volume fraction $\eta$ of monodisperse spheres is related
to the number density by $\eta=(\pi/6)\rho\sigma^3$, and
its maximum value is $\eta_{cp}=\pi/(3\sqrt{2})$, which is achieved
at close packing.

In order to generalise Eq. (\ref{asmono}) to a solid of polydisperse
spheres we have to make a further approximation. We neglect any
correlations between the size of spheres at adjacent lattice points.
In reality we would expect that on average smaller spheres will tend
to cluster around large spheres to reduce the strain on the lattice.
Every sphere is assumed to be in a cell formed from spheres
of the average diameter ${\overline\sigma}$. Then the cell
of a sphere of size $\sigma$ is of radius
$(a-(\sigma+{\overline\sigma})/2)$. This gives us
the free energy of a sphere of size $\sigma$; we simply
integrate over all sizes of spheres, weighting with $x(\sigma)$
to obtain the free energy per particle
\begin{eqnarray}
\frac{f_s}{kT}&=&-\ln\left(4\pi/3\right)-
3\int
x(\sigma)\ln\left[a-(\sigma+{\overline\sigma})/2\right]
{\rm d}\sigma
-\ln w
\nonumber\\
&=&-\ln\left(4\pi/3\right)+
3\left[1+
\frac{2}{w}\left(a-{\overline\sigma}-w/4\right)
\ln\left(a-{\overline\sigma}-w/4\right)
\right.\nonumber\\ &&-\left.
\frac{2}{w}\left(a-{\overline\sigma}+w/4\right)
\ln\left(a-{\overline\sigma}+w/4\right)\right]-\ln w,
\label{aspoly}
\end{eqnarray}
where the last term is the ideal mixing entropy \cite{salacuse82}
for our polydisperse
mixture \cite{salacuse82,gualtieri82}.
The lattice constant $a$ is still derived from
$a=(\eta_{cp}/\eta)^{1/3}$ with $\eta_{cp}$ for a monodisperse solid.
For polydisperse spheres with a distribution Eq. (\ref{xs})
the volume fraction is related to the total
number density by $\eta=(\pi/6)\rho(1+w^2/4)$.
For the free energy of the fluid phase we use the polydisperse
generalisation of Salacuse and Stell \cite{salacuse82} of the free energy of
Boublik, and Mansoori, Carnahan, Starling and Leland (BMCSL)
\cite{boublik70,mansoori71}.

Finding phase coexistence for polydisperse mixtures is very involved
\cite{barrat86,mcrae88,salacuse82,gualtieri82}, but see Refs.
\cite{sollich98,warren98}. Determining whether the fluid or the solid phase is
more stable, has the lower free energy, at a given density
is however straightforward, so we will
content ourselves with this.
In Fig. 1 we have plotted the free energies
of the fluid and solid phases with a polydispersity
$w/{\overline\sigma}=0.3$ ($s=0.087$).
The free energy of the solid phase is always
higher than that of the fluid phase and so is never the equilibrium phase.
This holds if and only if the polydispersity is greater than
$w=0.28$ ($s>0.082$).
However, the free energy of a solid phase with half the width,
here $w/{\overline\sigma}=0.15$
does drop below that of the fluid phase.
As the fluid has a higher free energy than solid phases with half
the polydispersity it is unstable with respect to them.
The phase separation will not of course be that complete
or simple. Describing the two phases as one with all the spheres
smaller than ${\overline\sigma}$ and the other with the spheres
larger than ${\overline\sigma}$ is a gross oversimplification.
However, we have proven that the fluid becomes unstable with respect
to two solid phases, whilst being stable with respect to one.
We have also plotted the free energy of the solid phase with
$w/{\overline\sigma}=0.1$. In the range of densities plotted in
Fig. 1 it is always higher than for $w/{\overline\sigma}=0.15$
so we expect that on crystallisation phase separation into two not
three solid phases will occur. Also, the free energy of the fluid
phase with $w/{\overline\sigma}=0.3$ is lower than that of the
fluid phase for either $w/{\overline\sigma}=0.15$ or 0.1.
As an aside, we note that
polydisperse sticky spheres also phase separate into
more than one solid phase \cite{sear}.

The free energy of phase
separated solid phases only drops below that of the fluid phase above
a volume fraction $\eta=0.6$.
Experiments on near-hard-sphere colloids with a polydispersity
$s\approx 0.04$ \cite{underwood94} show a glass transition
at $\eta=0.58$. Although of course we have not been able to find the
coexisting fluid and solid phases this at least suggests that, in
experiment, the combined phase-separation--crystallisation
phase transition may be prevented by a glass transition.
Even if it is not, the dynamics of a combined phase separation
and crystallisation will be slow \cite{pusey87,henderson98}.

Hard spheres with a polydispersity width $w>0.28$
($s>0.082$) cannot crystallise
into a single solid phase, its free energy is always higher than that of
the fluid. However, they can crystallise into two solid phases, we have
verified this for polydispersities up to $w=0.4$ ($s=0.12$).
We have not performed calculations beyond this because our solid phase
free energy may not be reliable at these polydispersities, and
because the density at which the fluid and solid free energies are equal
are so high ($\eta>0.63$) at these polydispersities that the BMCSL
free energy of the fluid phase may no longer be reliable.

In conclusion, experiments on near-hard-sphere colloids only find
crystallisation when
the polydispersity is no more than
approximately $s=0.075$ \cite{pusey87,pusey91}.
This is not the equilibrium behaviour of polydisperse hard spheres.
At equilibrium,
as the polydispersity is increased a second solid phase appears;
we predict that this occurs at a polydispersity $s<0.082$.
Each of the two solid phases is about half as polydisperse as the
original fluid.
For polydispersities $s$ of tens of per cent the picture
is much less clear:
there will now be spheres with very different diameters present which
may favour the formation of lattice types other than
a substitutionally-disordered close-packed lattice.
The formation of very different lattices has been observed
for binary mixtures of hard spheres
\cite{bartlett92,eldridge93}.
Another possibility is that highly polydisperse hard spheres do not
have an equilibrium solid phase.

It is a pleasure to thank J. Cuesta, D. Frenkel and P. Warren for
useful discussions.

%\newpage

%\newpage

\noindent
{\bf Fig. 1.}~~~
The free energy of polydisperse hard spheres in the fluid and solid
phase for several polydispersities. The solid curve is for
the fluid phase with a polydispersity width $w=0.3$. The
dotted, dashed and dot-dashed curves are for the solid phase
with polydispersity widths $w=0.3$, 0.15 and 0.1, respectively.
The dotted curve has been cut off just before a rapid divergence.

\end{multicols}

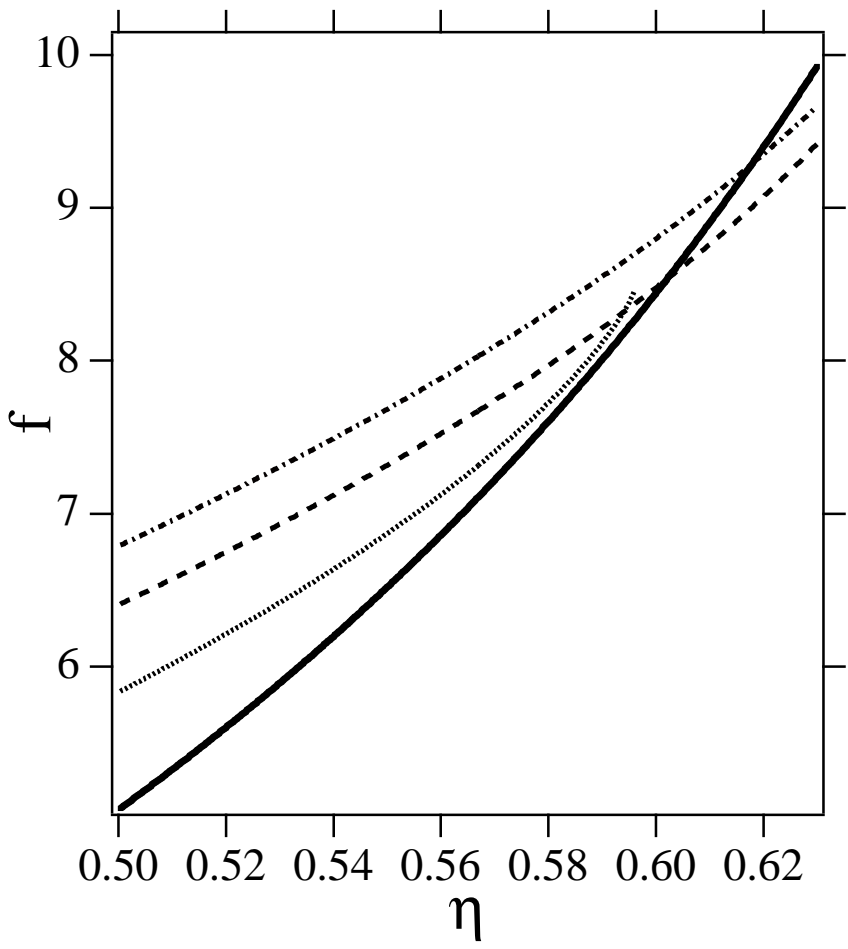
\end{document}